# Electronically Excited States of Vitamin $B_{12}$: Benchmark Calculations Including Time-Dependent Density Functional Theory and Correlated *Ab Initio* Methods

*Karina Kornobis[1], Neeraj Kumar[1], Bryan M. Wong[2], Piotr Lodowski[3], Maria Jaworska[3], Tadeusz Andruniów[4], Kenneth Rudd[5] and Pawel M. Kozlowski[1*]*

[1] Department of Chemistry, University of Louisville, Louisville, Kentucky 40292, USA,

[2] Materials Chemistry Department, Sandia National Laboratories, Livermore, California 94551, USA,

[3] Department of Theoretical Chemistry, Institute of Chemistry,

University of Silesia, Szkolna 9, PL-40 006 Katowice, Poland,

[4] Institute of Physical and Theoretical Chemistry, Department of Chemistry,

Wroclaw University of Technology, 50-370 Wroclaw, Poland,

[5] Centre for Theoretical and Computational Chemistry, Department of Chemistry,

University of Tromsø, 9037 Tromsø, Norway,

[*]Corresponding author, Phone: (502) 852-6609, Fax: (502) 852-8149. E-mail: pawel@louisville.edu

## ABSTRACT

Time-dependent density functional theory (TD-DFT) and correlated *ab initio* methods have been applied to the electronically excited states of vitamin $B_{12}$ (cyanocobalamin or CNCbl). Different experimental techniques have been used to probe the excited states of CNCbl, revealing many issues that remain poorly understood from an electronic structure point of view. Due to its efficient scaling with size, TD-DFT emerges as one of the most practical tools that can be used to predict the electronic properties of these fairly complex molecules. However, the description of excited states is strongly dependent on the type of functional used in the calculations. In the present contribution, the choice of a proper functional for vitamin $B_{12}$ was evaluated in terms of its agreement with both experimental results and correlated *ab initio* calculations. Three different functionals, i.e. B3LYP, BP86, and LC-BLYP, were tested. In addition, the effect of relative contributions of DFT and HF to the exchange-correlation functional was investigated as a function of the range-separation parameter, μ. The issues related to the underestimation of charge transfer (CT) excitation energies by TD-DFT was validated by Tozer's $\Lambda$ diagnostic, which measures the spatial overlap between occupied and virtual orbitals involved in the particular excitation. The nature of low-lying excited states was also analyzed based on a comparison of TD-DFT and *ab initio* results. Based on an extensive comparision against experimental results and *ab initio* benchmark calculations, the BP86 functional was found to be the most appropriate in describing the electronic properties of CNCbl. Finally, an analysis of electronic transitions and a new re-assignment of some excitations are discussed.

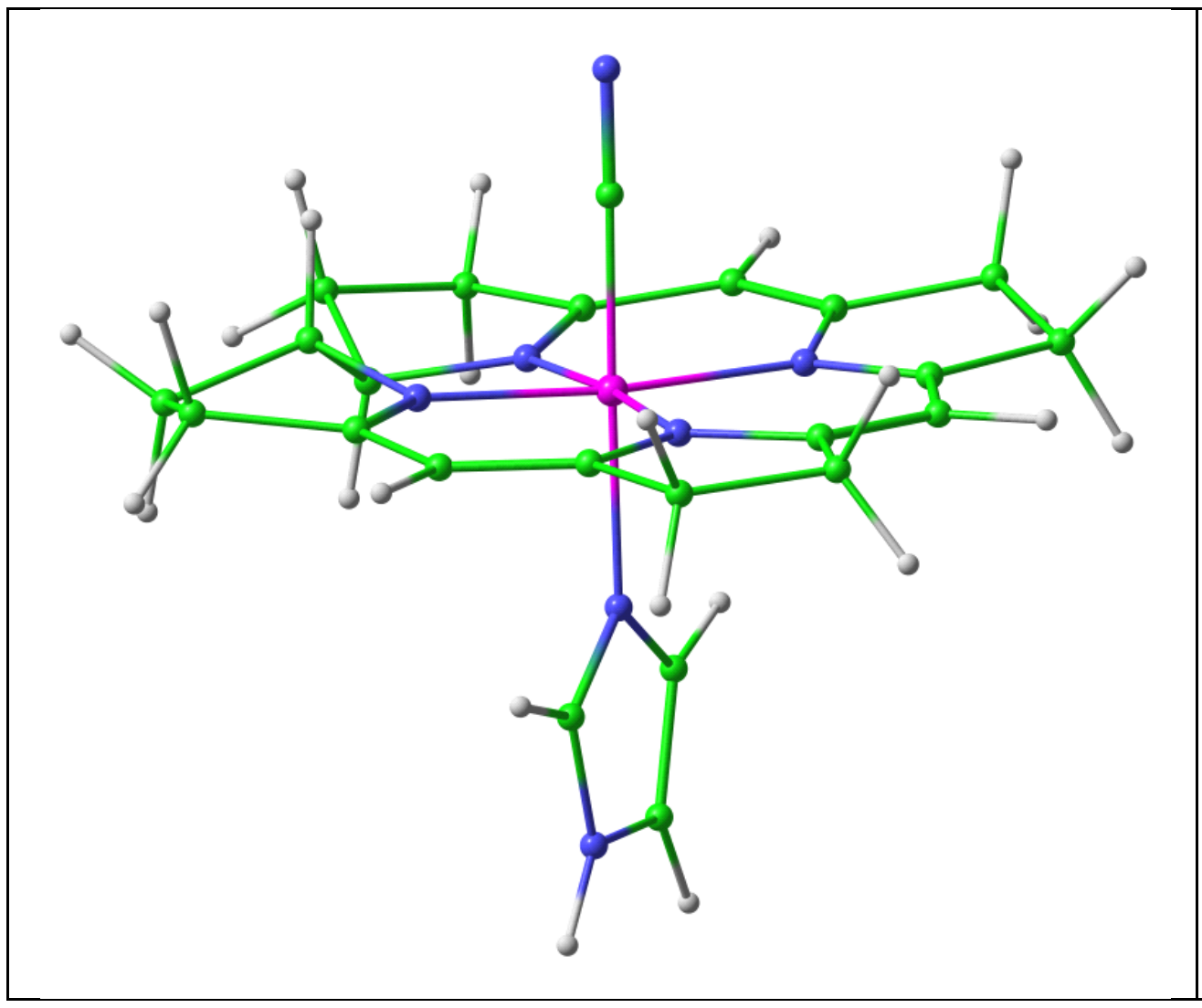

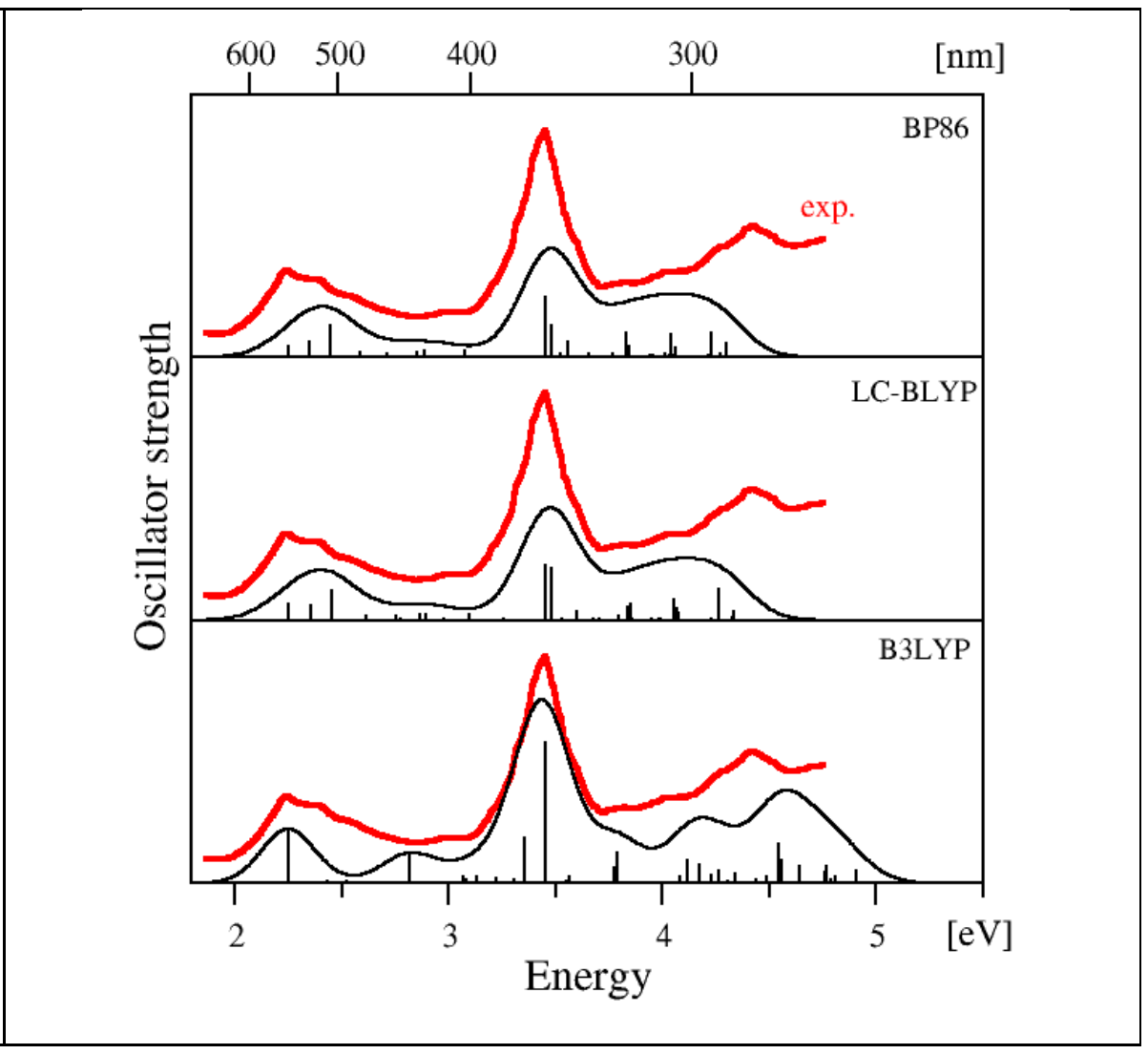

## 1. Introduction

Since its initial formulation,[1] linear-response time-dependent density functional theory (TD-DFT)[2,3] has become one of the most widely used computational tools to study electronically excited states of complex molecules.[4,5,6] It is generally accepted (at least for medium to large systems up to 300 second-row atoms) that certain electronic excitations can be accurately described within the TD-DFT framework. For generalized gradient approximation (GGA)[7,8] or hybrid exchange-correlation functionals,[9] computed excitation energies are often accurate to within a few tenths of an electronvolt. This applies to electronic transitions involving minimal charge transfer that lie well below the ionization potentials. However, for excitations to Rydberg states or for charge-transfer (CT) states, a significant underestimation of computed energies has been noticed.[10,11,12,13,14,15,16,17] While the former can be corrected using an asymptotic correction, the latter are more problematic mainly due to errors associated with self-interaction.[18,19]

CT-type excitation energies are often significantly underestimated by TD-DFT calculations employing conventional exchange-correlation functionals. In order to partially correct for this shortcoming, range-separated exchange-correlation functionals, such as Coulomb-attenuated B3LYP (CAM-B3LYP)[20] or long-range corrected BLYP (LC-BLYP),[21] have been designed for accurately calculating CT excitations in TD-DFT calculations. Nevertheless, the view that CT excitations are significantly underestimated by TD-DFT calculations needs to be interpreted with caution since these errors typically refer to long-range excitations with a low degree of spatial overlap between the occupied and virtual orbitals involved in the excitation. In order to judge the reliability of excitation energies, a simple test (referred to as the $\Lambda$ diagnostic)[22] has been suggested. For a given excitation, the degree of overlap is calculated using this $\Lambda$ diagnostic, which involves the inner product of the moduli of individual occupied and virtual Kohn-Sham orbitals. Its numerical value can range from 0 to 1, with values near zero signifying long-range excitation, while values near unity signifying a short–range, localized excitation. Based on its definition, excitation energies are significantly underestimated when $\Lambda$ is very small. It has further been proposed that an excitation with $\Lambda < 0.4$ from a GGA or $\Lambda < 0.3$ from a hybrid functional is likely to be significantly underestimated.[23,24] However, a high $\Lambda$ value does not guarantee that the excitation is correctly described by TD-DFT.

The use of range separated exchange-correlation functionals, such as CAM-B3LYP, does not necessarily guarantee that results are improved over conventional functionals either. In fact, the predicted spectra may worsen, as has been recently found for two cobalamins: cyanocobalamin (CNCbl or vitamin $B_{12}$) and methylcobalamin (MeCbl), where absorption (Abs), circular dichroism (CD) and magnetic CD (MCD) spectra based on TD-DFT calculations were simulated for direct comparison with experiment.[25] For these two bioinorganic complexes, the conventional BP86 functional performed better than CAM-B3LYP. It should also be noted that the $\Lambda$ diagnostic was computed for each excited state under consideration, and all computed $\Lambda$ values were greater than 0.4, indicating that the TD-DFT calculations were not affected by CT failure. The unexpected poor performance of the CAM-B3LYP functional may indicate that the inclusion of long-range Hartree-Fock exchange into TD-DFT overestimates the energy gap between the occupied and virtual orbitals, thus making the virtual-occupied energy difference a poor estimation for excitations energies. This type of sensitivity with respect to treatment of exact exchange seems to play a more important role in systems with transition metals rather than in organic molecules.[26,27]

In the present contribution, we continue our focus on vitamin $B_{12}$ (Figure 1) as one of the most important vital bioinorganic complexes. Electronically excited states of vitamin $B_{12}$ derivatives have been extensively investigated employing a variety of experimental techniques.[28,29,30,31,32,33] However, their excited-state properties still remain poorly understood from an electronic structure point of view. Since the TD-DFT framework is currently the only practical tool that can be used to routinely study excited states of these complex systems, the question naturally arises on how to select the appropriate functional for these studies. To some extent, this depends on the kind of spectroscopic properties one intends to investigate. If the target is the overall electronic spectrum and the objective is to assign certain electronic transitions, a number of excited states need to be computed to simulate the spectrum for direct comparison with experimental data. The performance of the functional is then judged based on how well the experimental spectrum is reproduced by the simulated one. However, when the target is an individual electronic state (or a manifold of low-lying excited states) the situation becomes somewhat more complicated. The performance based on a comparison with experiment is only possible if the identity of the particular electronic state is well established.[34] When an assignment is uncertain, the performance of the functional is analyzed by comparisons with excited state calculations carried out by high-level wavefunction-based methods. Typically CC2 or CASSCF/CASPT2 calculations are used to calibrate TD-DFT computed excited states.[35,36,37] In the present study, we discuss both our TD-DFT and *ab initio* calculations and then compare their accuracy in predicting vitamin $B_{12}$ excitation properties.

## 2. Computational Details

**2.1. Structural Models.** The full structure of vitamin $B_{12}$ was extracted from available high-resolution X-ray crystallographic data.[38] To decrease the computational cost associated with the complexity of the system (Figure 1), some simplifications to the initial structure were introduced. All side chains of the corrin ring as well as the nucleotide loop were truncated and replaced by hydrogen atoms. Previous studies have demonstrated that such simplified models of $B_{12}$ cofactors were appropriate for studying their structural and electronic properties.[39] The full structure of CNCbl also possesses dimethylbenzimidazole (DBI) as the lower axial base. In certain classes of $B_{12}$-dependent enzymes,[40] the DBI base is replaced by histidine (His) from the protein side chain. Hence, two structural models with different lower axial ligands were utilized in the present calculations. The first structural model contains imidazole (Im), and the second is axially ligated to DBI (Figure 2). Both models were optimized at the BP86/6-31G(d) level of theory using the Gaussian 09 suite of programs,[41] and their coordinates can be found in the Supporting Information (Tables S1 and S2). Taking into account that the main features observed in the electronic absorption spectra of cobalamins correspond to corrin $\pi \rightarrow \pi^*$ transitions, both models are appropriate to model electronically excited states of vitamin $B_{12}$.

**2.2. TD-DFT Calculations**. Several types of TD-DFT calculations have been carried out to explore the electronically excited states of CNCbl. To reproduce the electronic absorption spectrum of CNCbl, the 35 lowest excited states were calculated. All vertical excitation energies were computed at the BP86/6-31G(d) optimized ground state geometries described previously. Three different functionals, including BP86, B3LYP, and LC-BLYP, were utilized. Therange-separated LC-BLYP functional, which recovers the exact -1/*r* exchange dependence at large interelectronic distances, was calculated as a function of the range-separation parameter μ. Starting from μ = 0.00, which corresponds to a pure exchange-correlation density functional, the range-separation value was systematically increased up to 0.90 in increments of 0.05. The TD-DFT-based absorption spectra were analyzed in both gas phase and water solution environments, with the latter being modeled using a polarizable continuum model (PCM) and COnductor-like Screening MOdel (COSMO)[42]. The discussion about the nature of the low-lying excited states, however, focuses mainly on gas phase calculations which can be directly compared with high-level, correlated *ab initio* results. In certain instances, the electronically excited states were computed by employing the CAM-B3LYP functional. The TD-DFT calculations have been performed using the GAMESS,[43] TURBOMOLE,[44] and DALTON[45] programs.

**2.3. Correlated *ab initio* calculations.** Since one of the main objectives of the present study is an assessment of electronically excited states computed with different functionals, two types of *ab initio* calculations have been performed for benchmarking purposes. The CC2 and CASSCF computations have been carried out in order to determine the first four low-lying excited states. The former were performed with TURBOMOLE while latter were computed with PC-GAMESS/Firely QC package.[46] Due to a lack of dynamic correlation energy, the CASSCF calculations alone are not sufficiently accurate and need to be corrected by second order perturbation theory. A modified version of quasi-degenerate perturbation theory with multi-configurational self-consistent field reference function, referred to as MC-XQDPT2,[47] as implemented in the PC GAMESS/Firefly, was applied. In order to carry out the MC-XQDPT2 analysis, state average (SA) CASSCF calculations were first calculated using 12 electrons and 12 orbitals in the active space. It should be noted that QDPT2 correction mixes and alters the order of low-lying excited states based on initial CASSCF calculations. Thus, in order to obtain four lowest excited states at the CASSCF/MC-XQDPT2, it was required to initially compute 20 states at the SA-CASSCF level.

**2.4. Basis set.** In order to make a meaningful comparison between TD-DFT and *ab initio* calculations, the same basis set was used in both calculations. Taking into account the complexity of vitamin $B_{12}$, a double-zeta 6-31g(d) basis set was selected to perform the *ab initio* calculations. The TD-DFT results were re-evaluated with a larger TZVP triple zeta basis set. The structural model containing DBI as the lower axial ligand was also examined with a much larger aug-cc-pVDZ, basis set. It should be mentioned that the basis set quality significantly increases computational cost, and balancing accuracy against a reasonable computational time is crucial for the feasibility of the computations. Consequently, *ab initio* calculations with the triple-zeta quality basis set were not considered as a practical scheme, and these calculations were performed only with the 6-31G(d) basis set. We also note that since the CASSCF method largely considers static electronic correlation effects, and thus depends less on the nodal structure of the wave function, the 6-31G(d) basis will suffice for giving reasonable results for the MCSCF calculations. For the more dynamical correlation-oriented CC2 method, the 6-31G* basis may be less adequate, but we will see that there are other concerns regarding the reliability of the CC2 results.

## 3. Results and Discussion

The performance of DFT in modeling electronic and structural properties associated with the ground state of $B_{12}$ cofactors has been the subject of several studies.[48,49] It was concluded that GGA-type functionals (such as BP86) are particularly well suited for describing their ground state properties. The recent review by Jensen and Ryde[50] provides a more detailed discussion. Other properties such as absorption spectra, CD, and MCD data were also analyzed based on realistic corrin-based models. Results based on the BP86 functional were in much better agreement with experimental data than those based on CAM-B3LYP, demonstrating the improved performance of BP86 compared to hybrid or Coulomb-attenuated functionals.

**3.1. Analysis of absorption data.** The manifold of low-lying excited states was computed to cover the spectral range up to 250 nm (ca. 5 eV). The calculated 35 vertical excitations and oscillator strengths were used to simulate the absorption spectra for direct comparison with experiment (Figures 5 and 6) in both gas-phase and PCM-based water solvent model. In order to compare with the experimental spectrum,[51] each TD-DFT electronic transition was broadened with a Gaussian function having a width of $\Gamma$ = 0.174 eV. Since the LC-BLYP and BP86-simulated spectra were almost identical, we performed additional computations using the LC-BLYP functional as a function of the range-separation μ parameter (Figure S5, Supporting Information). By varying the values of μ, the influence of relative contributions of DFT and HF to the exchange-correlation energy on the spectra were assessed. The spectra corresponding to μ values ranging from 0.00 to 0.85 were simulated and directly compared with experiment (Figures S1 and S2, Supporting Information). The absorption profiles obtained in both environments (gas phase, and PCM-based water solution) showed the best agreement with experiment only for the first few low μ values, indicating the use of GGA functionals is a better choice in predicting electronic properties of CNCbl.

Previous studies on cobalt corrinoids have demonstrated that electronic excitations based on TD-DFT calculations utilizing hybrid functionals were systematically overestimated.[52] Since the overestimation was systematic, it could be compensated by scaling. Initially, a simple shift of electronic transitions was applied for models of vitamin $B_{12}$; however, for the alkylcobalamins, such as methylcobalamin (MeCbl) or adenosylcobalamin (AdoCbl), a better agreement with experiment was obtained when two scaling parameters were applied.[53,54] To achieve better correlation with experimental data, each electronic transition was scaled as $E^i_{scaled} = \xi E^i_{TD\text{-}DFT} + E_{shift}$, where $\xi$ and $E_{shift}$ are a specific set of parameters for a given functional and a corrinoid. The determination of the optimal pair of parameters was based on the assumption that electronic excitations associated with $\alpha$ and $\gamma$ bands should be reproduced accurately. The unscaled spectra along with scaling factors can be found in the Supporting Information (Figures S3 and S4, Tables S6 and S7)

The comparison of the simulated absorption spectra (Figure 5 and 6) indicates overall good agreement with experiment regardless of the functional chosen in TD-DFT calculations. However, the detailed analysis of electronic transitions reveals some differences. Whereas the high energy $\gamma$ band is in accordance with experiment and gains the largest intensity in the B3LYP-simulated spectrum, the oscillator strengths of the same band predicted by the remaining two methods are significantly lowered. Moreover, in contrast to B3LYP results, the BP86 and LC-BLYP gas phase data indicate the involvement of two (instead of one) intense electronic transitions. Based only on the reproduction of the experimental $\gamma$ band in gas phase, one can conclude the better suitability of the B3LYP functional in simulating absorption spectra of CNCbl. At first glance, this may imply that different functionals are good for different parts of the spectrum. However, it is important to notice that the change of the environment improves the intensity pattern in the $\gamma$ region for all three functionals and reveals the presence of a single electronic transition regardless of the method applied. Moreover, the $\alpha/\beta$ band seems to be better reproduced by BP86 and LC-BLYP functionals. The interpretation of simulated CD and MCD spectra implied the existence of a manifold of electronic excitations in the $\alpha/\beta$ part of the CNCbl absorption spectrum. The data shown in the present study indicates that there are multiple excitations present in the low-energy part of the spectra only if BP86 or LC-BLYP is applied. Both gas phase as well as PCM data (Figure 5), obtained from BP86 and LC-BLYP calculations, confirm such an assignment. In contrast, the $\alpha/\beta$ band in B3LYP-computed profiles is composed of one intense excitation regardless of the environment (Figure 5). These results are more in line with earlier proposals, suggesting vibrational progression of a single electronic transition in the α/β region of CNCbl spectrum. As can be noticed, the choice of the density functional is crucial in terms of a reliable description of particular bands in CNCbl absorption spectra. However, the final statement about the correct applicability of the particular functional demands a detailed analysis of the nature of the excited states coupled with a reasonable evaluation of the theoretical methods in terms of their correspondence to experimental results. Taking into account the photochemistry of $B_{12}$ cofactors, the most important salient features arise from the examination of the first few low-lying electronic excited states. Consequently, the comparison of TD-DFT results with correlated *ab initio* data (presented further in this paper) allows for a reliable justification of the suitability of a particular density functional to study electronic properties of CNCbl.

**3.2. Lambda diagnostic for low-lying excited states.** Up to this point, the performance of different functionals, including GGA and hybrid, has been discussed in the context of how well they reproduce absorption spectra. Although functionals analyzed in the present study tend to provide satisfactory agreement with experiment, the CT-type excitations are often very poorly described by TD-DFT. In order to justify the reliability of vertical excitations of CNCbl predicted at the TD-DFT framework with different functionals, calculations involving the $\Lambda$ diagnostic were performed. Figure 3 displays the $\Lambda$ values corresponding to the manifold of low-lying excited states of the vitamin $B_{12}$ model having imidazole as the axial base, i.e. Im-[$Co^{III}$(corrin)]-$CN^{+}$ (Figure 2, upper panel). As can be seen already with the 6-31g(d) basis set, none of the computed electronic transitions have noticeably low $\Lambda$ values, indicating that our calculations do not suffer from the CT problem. We have found that an increase in the basis set does not dramatically influence the $\Lambda$ values. On the other hand, when the same calculations were carried out for the model having DBI as the axial base, i.e. DBI-[$Co^{III}$(corrin)]-$CN^{+}$ (Figure 2, lower panel) the problem related to the CT issue was more noticeable (Figure 4). It appears that this model is more susceptible to CT problems which is due to the larger lower axial base having more extensive aromatic character and a longer Co-N(axial) (2.09 Å as compared to 2.05 Å for Im-[$Co^{III}$(corrin)]-$CN^{+}$ model) distance. According to the $\Lambda$ diagnostic, the predictions from the BP86/6-31G(d) calculations indicate that the $S_4$ and $S_5$ states have CT character since their $\Lambda$ values are lower than 0.4. The problem disappears with an increase in basis set size, although this is not due to a change in character induced by the change of the basis set, but rather the improved ability of the more diffuse aug-cc-pVDZ basis set to model the rather short-ranged CT of this excitation. The results based on calculations with the TZVP basis set show improvement for $S_4$ and $S_5$ states but also simultaneously decrease the $\Lambda$ values for $S_3$ and $S_6$. At the same time, the calculations for low-lying excited states of DBI-[$Co^{III}$(corrin)]-$CN^+$, employing the aug-cc-pVDZ basis set, do not produce excitation energies suffering with the CT failure (Figure 4). However, it is important to note that the size of basis set doubles (from ~600 to ~1200), making such calculations feasible but not practical.

The conclusion coming from our analysis indicates that the CNCbl excitation energies calculated using TD-DFT are not affected by CT failure. The lower $\Lambda$ values in the case of DBI-[$Co^{III}$(corrin)]-$CN^+$ model are related to the quality of the basis set and can not be a reason for disregarding TD-DFT/BP86-based results. Moreover, the increased basis set does not change the character of the excitations, allowing for a reasonable comparison of the nature of low-lying transitions predicted by TD-DFT and *ab initio* methods employing the 6-31G(d) basis set.

**3.3. Comparison with *ab initio* calculations.** We now turn our attention to *ab initio* methods in order to analyze in detail the nature of low-lying electronic transitions of CNCbl. For medium-sized systems, the CC2 method is commonly used to compute excitation energies to serve as accurate benchmarks for TD-DFT calculations. We initially focused on the four lowest states of CNCbl in order to estimate their energies at the CC2 level of theory. However, as shown in Table 1, the CC2 excitation energies are severely underestimated due to a significant multi-reference character indicated by the D1 diagnostic.[55] The D1 diagnostic essentially measures the quality of the HF reference wavefunction, with values less than 0.05 indicating that a single determinant is appropriate. We found that all of the 4 lowest electronic states of CNCbl were significantly above the 0.05 threshold, demonstrating a severe deficiency in the HF reference wavefunction for these metal-containing complexes.

Consequently, CASSCF calculations were applied as a more appropriate method for studying electronically excited states of CNCbl. The CASSCF approach is very challenging when applied to molecules containing transition metals, and the results strongly depend on the choice of active space. Since CNCbl contains a cobalt atom in its structure, the proper selection of active space enforces the use of the metal's *d* orbitals. Also, a complementary set of 3*d* orbitals has to be added to the active space in order to take into account double-shell effect.[56] When the complex contains a tetrapyrrolic ligand (such as corrin) in addition to cobalt *d* orbitals, the proper active space also requires inclusion of $\pi$ orbitals of corrin as well as $\sigma$ orbitals associated with axial bonding. Keeping in mind these constraints, the active space was initially selected using orbitals previously applied in CASSCF calculations for cob(I)alamin[57] as well as corrole[58] systems. Several possible compositions have been tested, leading to the conclusion that the most appropriate choice was an active space based on the distribution of 12 electrons between 12 orbitals. More specifically, five *d* Co and one $\pi$ corrin occupied MOs were combined with the corresponding virtual orbitals, and their isosurface plots are presented on Figure 7.

The analysis of vertical excitations predicted at the CASSCF(12,12) level of theory shows an overestimation of the energies in comparison to those computed at the TD-DFT level with different functionals (Table 1). The inspection of the nature of the energetically lowest CASSCF(12,12) excitations (Table S8, Supporting Information) revealed a domination of $d \rightarrow d$ transitions, contrary to all other results (Table 3). This finding also contradicts the expectation that electronic excitations possessing non-zero transition dipole moments mainly involve corrin $\pi \rightarrow \pi^*$ transitions. In order to obtain reliable results, second-order perturbation theory was applied using the MC-XQDPT2 scheme based on the SA CASSCF wavefunction. Inclusion of dynamical correlation energy changes the nature of the electronic transitions, and more states having mixed $d/\pi \rightarrow \pi^*$ character are produced (Table 2). The specific mixing is shown in Figure 8, where only the four lowest MC-XQDPT2 excited states were depicted since those are the only ones considered for comparison with the TDDFT results. None of the first few electronic excitations predicted at the MC-XQDPT2 level retain the same character as the CASSCF states. In addition, the involvement of dynamic correlation varies for different states and becomes more significant in states having $d \rightarrow \pi^*$ and $\pi \rightarrow \pi^*$ transitions.

To further explore the nature of low-lying excited states of CNCbl, we compared the results obtained from TD-DFT and MC-XQDPT2 calculations (Tables 3 and 2). The electronic absorption spectra of cobalamins (at least the most intense transitions) are dominated by corrin $\pi \rightarrow \pi^*$ excitations. However, as noted before, experiments focusing on the photochemistry of $B_{12}$ cofactors suggest the possibility of CT character for the energetically lowest excitations. Data presented in Table 2 confirm such predictions as none of the MC-XQDPT2-calculated states can be assigned as having a pure nature. In fact, the $S_1$, $S_3$, and $S_4$ states are of mixed $d \rightarrow \pi^*$ and $\pi \rightarrow \pi^*$ character, whereas $S_2$ is primarily a $d \rightarrow \pi^*$ transition. Based on this assignment and the TD-DFT results presented in Table 3, it can be concluded that hybrid functionals fail in predicting the correct nature of the electronic excited states of CNCbl. On the other hand, the excitations computed by BP86 show good agreement with MC-XQDPT2 calculations. The $S_1$ and $S_3$ states are composed of mixed $d/\pi \rightarrow \pi^*$ and $\pi \rightarrow \pi^*$ transitions, and $S_2$ is dominated by $d/\pi \rightarrow \pi^*$. Similar results are obtained from the LC-BLYP functional when the range-separation parameter $\mu$ is zero (which essentially reduces the LC-BLYP formalism to a GGA-type functional). However, in order to evaluate the impact of increasing the amount of exact HF exchange on the character of low-lying electronic transitions, the performance of the LC-BLYP method was analyzed by varying the value of $\mu$ from 0.00 to 0.90 (Table S9, Supporting Information). The description of low-lying excited states of CNCbl worsens with an increased value of long-range HF exchange as indicated by a larger value of $\mu$. Therefore, we conclude that the most reliable description of CNCbl excitations is predicted by BP86 calculations which imply the involvement of corrin $\pi \rightarrow \pi^*$, as well as charge transfer transitions involving cobalt $d$ orbitals.

**3.4. Assignment of Electronic Excitations.** The previous studies employing the B3LYP functional lead to the assignment of the majority of bands observed in the absorption spectrum of CNCbl.[30,52] As in other corrinoid species, the B3LYP calculations yield a much more blue-shifted transition with respect to experimental values. Although the spectral shifts can be compensated by scaling, other problems with the B3LYP functional such as the underestimation of Co-C bond dissociation energy, overestimation of Co-$N_{axial}$ bond distance, as well as the description of the nature of low-lying excited states can not be disregarded. Hence, we focus our discussion on our TD-DFT/BP86 calculations in order to present a re-assignment of electronic transitions of CNCbl.

The description of the singlet excited states of CNCbl is presented in Table 4 and 5. The longest wavelength part of the CNCbl absorption spectrum ($\alpha/\beta$ region) shows several bands at 550, 517, 485sh (sh = shoulder) nm, that were originally interpreted as a vibrational progression of one intense $\pi \rightarrow \pi^*$ electronic transition, giving rise to the corrin –C=C– stretching vibrations. However, according to previous findings based on CD and MCD spectra, it is not necessary to make such an assumption in order to understand the absorption spectrum of CNCbl. The lowest energy singlet states $S_1$ - $S_{10}$ are transitions involving several highest occupied MOs and lowest unoccupied MOs (Figure 9). As described in the previous section, the first three of them are dominated by both d/$\pi \rightarrow \pi^*$ and $\pi \rightarrow \pi^*$ transitions. In addition, the $S_3$ state at 474 nm also involves a 14% contribution from a HOMO to LUMO+1 transition that corresponds to a $\pi \rightarrow$ d transition. It is interesting to note that the $S_4$ transition at 455 nm is primarily comprised of a $\pi \rightarrow$ d excitation and may also be included in the α/β band. In fact, experiments based on ultrafast transient absorption spectroscopy have indicated the significant role of CTs in the photolysis of cobalamin species. In particular, the metastable $S_1$ state of CNCbl was found to have a ligand-to-metal charge transfer (LMCT) character. Although the BP86-computed results suggest contributions to only the third or fourth state, their presence should be interpreted with caution, especially since our COSMO results (Table 5) show very high 40% and 53% contribution of $\pi \rightarrow d$ transition to $S_2$ and $S_3$ excitations, respectively.

The further part of the low-energy region of CNCbl spectrum reveals excitations with very small intensities. Among them, only two possess ~0.01 oscillator strengths. The first low-oscillator strength transition involves the $S_8$ state, which can be characterized as a $\pi \rightarrow \sigma^*$ excitation. The same transition in the simulated solution spectra is shifted to a lower energy region and found as a fifth excited state. The second intensity corresponds to the $S_{10}$ state, calculated in gas phase, and is composed mainly of $d/\pi \rightarrow \pi^*$ transition. Nevertheless, two additional states, $S_{11}$ and $S_{12}$ calculated in solution, possess almost identical energies to the $S_{10}$ state, with the $S_{12}$ having the highest intensity. Although $S_{12}$ reveals a large, 41% contribution from a $\pi \rightarrow \pi^*$ transition, it is similar to the remaining two excitations which are also composed of mixed $d/\pi \rightarrow \pi^*$ transitions. Taking into account only the intensity pattern, two gas phase excitations ($S_8$ and $S_{10}$) could be attributed to the D and E bands in the CNCbl absorption spectrum. Previous CD/MCD reports, however, suggested that such estimations have to be done very carefully as some of the excitations, especially those including $d \rightarrow d$ transitions, might not obtain sufficient intensities in calculated absorbance profiles. In fact, in the analyzed region of the spectrum (first 12 excited states) there are several other electronic transitions that can be described as $d/\pi \rightarrow d$, $d/\pi \rightarrow \pi^*$, $d/\pi \rightarrow \sigma^*$ and $\pi \rightarrow \pi^*$ and should not be excluded unambiguously from the assignment of the D/E band.

At higher energies, the $\gamma$ part of the spectrum reveals absorption features at 360, 322, and 306 nm. Two calculated transitions of the highest intensities in that region are found at 350 and 348 nm. The former has an oscillator strength more than twice as large as the latter and is mainly a $\pi \rightarrow \pi^*$ transition. At the same time, the excitation at 348 nm includes some contribution from $\pi$ orbitals localized on the axial cyano group, leading to a mixed $\pi_{CN}/\pi \rightarrow \pi^*$ nature of the discussed transition. As mentioned in section 3.1, the change of the environment leads to the presence of a single electronic transition corresponding to the $\gamma$ part of the spectrum. A detailed analysis of the BP86 transitions calculated in the presence of a solvent model (Table 5) indicates that the most intense peak corresponds to the $S_{17}$ state possessing 32, 29, and 12% contributions from $d_{yz}/\pi/\pi_{CN} \rightarrow \pi^*$, $\pi \rightarrow \pi^*$ and $d_{xz}/\pi/\pi_{Im} \rightarrow \pi^*$, respectively. This observation leads to the conclusion that the $\gamma$ band of CNCbl is not necessarily a pure $\pi \rightarrow \pi^*$ electronic transition. When describing the nature of this excitation, one has to consider the involvement of orbitals localized on the corrin ring ($\pi$), as well as axial bonds of the system (d, $\pi_{CN}$) as indicated by both *in vacuo* and COSMO data.

Together with the increase in energy, two additional spectral features, at 322 and 306 nm have been extracted from the experimental absorption profiles of CNCbl. However, the unambiguous assignment of these bands is hampered by the congestion of calculated electronic transitions in the high-energy part of the spectra, which have very small oscillator strengths. Following the intensity pattern, the band at 322 nm can be attributed to $S_{23}$ obtained from calculations in gas phase and described as a $d/\pi \rightarrow \pi^*$ transition. The same state in solution, however, is blue shifted by ~7 nm, and its oscillator strength decreases significantly. At the same time, the following two excitations ($S_{24}$ and $S_{25}$) in solution gain almost similar intensities, making them plausible contributors to the experimentally observed band. However, their nature is mixed due to the involvement of several occupied and unoccupied MOs (Table 5). The transitions are composed not only of $d/\pi/\pi_{CN} \rightarrow d$ and $d/\pi/\pi_{CN} \rightarrow \pi^*$ excitations but also of $d/\pi/\pi_{CN} \rightarrow \sigma^*(d_{z^2})$, ($S_{24}$) and $\pi \rightarrow \pi^*$, ($S_{25}$), making the assignment of the experimental band (at 322 nm) a very difficult task. Assignment of the experimental band at 306 nm is even less straightforward as there are a number of excitations having comparable oscillator strengths, in the calculated spectral range from ~303 to ~290 nm in gas phase. In addition, most of them possess small contributions from several electronic transitions and, hence, none of them has a pure character. The most intense calculated transition ascribed to the 306 nm experimental band is the transition at 289 nm. This transition is comprised of $\pi_{CN}/\pi \rightarrow \sigma^*_{Co\text{-}CN}$ and $\sigma_{Co\text{-}CN} \rightarrow d$ excitations.

The inclusion of solvent lowers the oscillator strengths of the particular excitations in this region. The most intense excitation corresponds to the $S_{32}$ state which refers to a $\pi/\pi_{CN} \rightarrow \pi^*$ transition. Taking into account only the intensity, this transition might be attributed to the experimental band at 306 nm. However, its significant (~25 nm) blue shift as compared to experiment, as well as the lack of a similar counterpart obtained from calculations in the gas phase, makes such an assignment speculative.

In the energy range of the $\delta$ band, there are several calculated transitions of relatively large oscillator strengths 277, 275, 266, and 265 nm. The former two transitions are $d/\pi_{CN}/\pi \rightarrow \pi^*_{Im}$ and $\pi_{CN}/\pi \rightarrow \pi^*$ excitations. The transition at 266 nm involves excitations of $\sigma_{Corr}/\pi_{Im} \rightarrow \pi^*$ and $\pi/d/\pi_{CN} \rightarrow \pi^*$ character, and the transition at 265 nm is composed of $\sigma_{Corr}/\pi_{Im} \rightarrow \pi^*$ and $\pi/d/\pi_{CN} \rightarrow d$ excitations. Similarly to the lower-energy region containing bands at 322 and 306 nm, the change of the environment decreased the oscillator strengths of the individual excitations contributing to the $\delta$ band as well. In addition, the calculated transitions change their character, making the final assignment of this region in the absorption spectrum of CNCbl very counterintuitive. The most intense band in this spectral part, resulting from calculations in solution (Table 5), correspond to the $S_{40}$ state which is composed of $\pi_{Im}/\pi_{CN} \rightarrow d$ and $\pi_{CN}/\pi \rightarrow d$ transitions. Its presence, at 260 nm, suggests a possible correspondence to the experimental band at 264 nm but in order to support or refute this proposal, further studies are required including both experimental and theoretical investigations.

## 4. Summary and Conclusions

Although electronically excited states of vitamin $B_{12}$ derivatives have been probed employing a variety of experimental techniques, their exact nature remains poorly understood from an electronic structure point of view. In order to shed light on their complex spectra, the TD-DFT framework offers a practical method that can be applied to predict the excited states of complex bioinorganic molecules such as vitamin $B_{12}$. Nevertheless, a critical step in TD-DFT calculations is the choice of an appropriate functional which was the focal point of the present study.

In the present contribution, the choice of the proper functional for vitamin $B_{12}$ was evaluated by taking into account comparisons with experimental data as well as correlated *ab initio* wavefunction-based calculations. Three different methods received special attention: GGA (BP86), hybrid (B3LYP), and the range-separated (LC-BLYP) functionals, which was tested as a function of the damping parameter, μ. A comprehensive analysis revealed that only results based on BP86 and LC-BLYP, with μ being close to zero, are consistent with both experimental results and high-level *ab initio* calculations. The finding that a GGA-type functional gives a better description than a hybrid one may appear surprising; however, it also reflects the fact that the BP86 functional gives a reliable description of the energy gap between the occupied and virtual orbitals in organometallic complexes, thus making the virtual-occupied energy difference a good estimation for excitations energies. In order to assess contributions due to CT-type excitations, which are often found to be inaccurately predicted by TD-DFT calculations, we also presented extensive calculations of the$\Lambda$ diagnostic test. The results of our findings dispelled suspicions about possible CT problems and confirmed the legitimacy of TD-DFT in studying electronic properties of CNCbl. Consequently, the use of BP86 is suggested as an expedient and accurate choice in calculating electronically excited states of CNCbl.

The conclusion reached in the present study regarding the BP86 functional may have several important implications for computational studies involving modeling of cobalt corrinoids. First, it puts into question earlier studies where assignment of electronic transitions for vitamin $B_{12}$ have been proposed using the B3LYP functional. Although, traditionally, electronic transitions of $B_{12}$ cofactors have been assigned mostly as having $\pi \rightarrow \pi^*$ character, the BP86 results indicate that many excitations possess significant contributions from transitions involving cobalt d orbitals. Second, it further supports conclusions reached in our recent study, where CD and MCD data were analyzed, showing the advantage of GGA over hybrid functionals in modeling excited states of $B_{12}$ derivatives. This particularly applies to the lowest energy $\alpha/\beta$ band, where multiple electronic transitions are observed rather than the vibrational progression of a single electronic excitation. Third, from a theoretical perspective, targeting a particular electronically excited state becomes a moot case when there is a lack of experimental evidence supporting computed results. This particularly applies to photochemical and photophysical events where low-lying excited states play a significant role. It has been noted that in CNCbl these states are related to LMCT transitions. The examination of these excitations at different levels of theory presented in this study confirm the applicability of the BP86 functional in describing electronic properties of CNCbl due to their best correspondence to CASSCF/MC-XQDPT2 results. This conclusion may assist in prospective studies explaining the photochemical processes of CNCbl. Overall, the present study demonstrates progressive investigations of electronically excited states of CNCbl and a reliable evaluation of TD-DFT methods in studying the properties of $B_{12}$ cofactors.

## Acknowledgments

This work was supported by Ministry of Science and Higher Education (Poland) under grant No. N204 028336. The TURBOMOLE calculations were carried out in the Wrocław Centre for Networking and Supercomputing, WCSS, Wrocław, Poland, http://www.wcss.wroc.pl, under calculational Grant No. 51/96. Sandia is a multiprogram laboratory operated by Sandia Corporation, a Lockheed Martin Company, for the United States Department of Energy's National Nuclear Security Administration under contract DE-AC04-94AL85000.

**Table 1.** Vertical excitation energies (eV) of Im-[$Co^{III}$(corrin)]-$CN^+$ calculated for the first four low-lying states with different methods and the 6-31G(d) basis set.

| Method | $S_1$ | $S_2$ | $S_3$ | $S_4$ |
|---|---|---|---|---|
| TD-DFT | | | | |
| B3LYP | 2.75 | 2.83 | 2.93 | 3.10 |
| BP86 | 2.43 | 2.53 | 2.62 | 2.73 |
| LC-BLYP (μ=0.00) | 2.43 | 2.54 | 2.62 | 2.74 |
| LC-BLYP (μ=0.10) | 2.53 | 2.66 | 2.70 | 2.82 |
| LC-BLYP (μ=0.20) | 2.74 | 3.05 | 3.12 | 3.14 |
| LC-BLYP (μ=0.30) | 2.86 | 3.08 | 3.17 | 3.33 |
| *Ab initio* | | | | |
| CC2 | 1.80 | 2.27 | 2.57 | 2.63 |
| CASSCF(12,12) | 3.44 | 3.48 | 3.60 | 4.75 |
| CASSCF/MC-XQDPT2 | 2.08 | 2.32 | 2.68 | 2.98 |

**Table 2.** The first four low-lying excited states of Im-[Co$^{III}$(corrin)]-CN$^+$ obtained from CASSCF(12,12) and MC-XQDPT2 calculations. CSF stands for Configuration State Function.

| State | E(eV) | Coeff. of CAS state in MC/XQDPT2 | weight (%) of CAS state in MC-XQDPT2 | CAS state | Coeff. of CSF [%] | Character |
|---|---|---|---|---|---|---|
| $S_0$ | 0.00 | -0.924238 | 85 | 1 | | |
| $S_1$ | 2.08 | -0.601547 | 36 | 16 | 24 | $d_{yz} \rightarrow \pi^*$ |
| | | | | | 10 | $d_{xz} \rightarrow \pi^*$ |
| | | | | | 4 | $d_{xz}/\pi \rightarrow \pi^*/ \sigma^*(d_{z^2})$ |
| | | -0.482393 | 23 | 11 | 18 | $\pi \rightarrow \pi^*$ |
| | | | | | 12 | $\sigma(d_{z^2}) \rightarrow \pi^*$ |
| | | | | | 11 | $d_{xy} \rightarrow \pi^*$ |
| $S_2$ | 2.32 | -0.793131 | 63 | 14 | 31 | $d_{x^2-y^2} \rightarrow \pi^*$ |
| | | | | | 8 | $d_{xz} \rightarrow \pi^*$ |
| | | | | | 8 | $d_{x^2-y^2}/\sigma(d_{z^2}) \rightarrow \pi^*/\sigma^*(d_{z^2})$ |
| | | 0.530156 | 28 | 19 | 16 | $d_{x^2-y^2} \rightarrow \pi^*$ |
| | | | | | 15 | $d_{x^2-y^2}/\pi \rightarrow \pi^*/d_{xy}+\sigma^*$ |
| | | | | | 7 | $d_{xy}/d_{x^2-y^2} \rightarrow \pi^*/\sigma^*(d_{z^2})$ |
| $S_3$ | 2.68 | -0.751397 | 56 | 15 | 21 | $d_{xz} \rightarrow \pi^*$ |
| | | | | | 16 | $d_{yz} \rightarrow \pi^*$ |
| | | | | | 6 | $d_{x^2-y^2} \rightarrow \pi^*$ |
| | | | | | 6 | $d_{xy} \rightarrow \pi^*$ |
| | | -0.323601 | 10 | 11 | 18 | $\pi \rightarrow \pi^*$ |
| | | | | | 11 | $\sigma(d_{z^2}) \rightarrow \pi^*$ |
| | | | | | 11 | $d_{xy} \rightarrow \pi^*$□ |
| $S_4$ | 2.98 | 0.454615 | 21 | 16 | 24 | $d_{yz} \rightarrow \pi^*$ |
| | | | | | 10 | $d_{xz} \rightarrow \pi^*$ |
| | | | | | 4 | $d_{xz}/\pi \rightarrow \pi^*/ \sigma^*(d_{z^2})$ |
| | | -0.443262 | 20 | 11 | 18 | $\pi \rightarrow \pi^*$ |
| | | | | | 11 | $\sigma(d_{z^2}) \rightarrow \pi^*$ |
| | | | | | 11 | $d_{xy} \rightarrow \pi^*$ |

**Table 3.** Composition of low-lying electronically excited states of Im-[$Co^{III}$(corrin)]-$CN^+$ based on TD-DFT calculations with the use of different density functionals and the 6-31G(d) basis set.

| State | Functional | Energy (eV) | Character |
|---|---|---|---|
| $S_1$ | B3LYP | 2.75 | 79%($\pi \rightarrow \pi^*$) |
| | BP86 | 2.43 | 30%($\pi \rightarrow \pi^*$) + 15%($d_{xz}+\pi_{CN}+\pi \rightarrow \pi^*$) |
| | LC-BLYP (μ=0) | 2.43 | 72%($\pi \rightarrow \pi^*$) + 26%($d_{xz}+\pi \rightarrow \pi^*$) |
| $S_2$ | B3LYP | 2.83 | 22%($d_{xz}+\pi_{CN} \rightarrow \sigma^*+d_{xy}$) + 17%($d_{xz}+\pi \rightarrow \sigma^*+d_{xy}$) |
| | BP86 | 2.53 | 36%($d_{yz}+\pi \rightarrow \pi^*$) |
| | LC-BLYP (μ=0) | 2.54 | 83%($\pi+d_{yz} \rightarrow \pi^*$) |
| $S_3$ | B3LYP | 2.93 | 24%($\pi \rightarrow \sigma^*(d_{z^2})$) + 20%($d_{xz}+\pi_{CN} \rightarrow \sigma^*(d_{z^2})$) + 10%($\pi_{Im}+\pi_{CN} \rightarrow \sigma^*(d_{z2})$) |
| | BP86 | 2.62 | 28%($d_{xz}+\pi_{CN}+\pi \rightarrow \pi^*$) + 9%($\pi \rightarrow \pi^*$) + 7%($\pi \rightarrow d_{xy}+n$) |
| | LC-BLYP (μ=0) | 2.62 | 65%($d_{xz}+\pi \rightarrow \pi^*$) + 22%($\pi \rightarrow \pi^*$) |
| $S_4$ | B3LYP | 3.10 | 48%($d_{xy}+\sigma \rightarrow \sigma^*+d_{xy}$) + 16%($d_{xy}+\sigma \rightarrow \sigma^*(d_{z^2})$) + 10%($\pi+d_{xz} \rightarrow \sigma^*+d_{xy}$) |
| | BP86 | 2.73 | 37%($\pi \rightarrow d_{xy}+n$) |
| | LC-BLYP (μ=0) | 2.74 | 86%($\pi \rightarrow \sigma^*+d_{xy}$) |

**Table 4.** TD-DFT/BP86/6-31G(d)-based electronic transitions for the singlet states of Im-[$Co^{III}$(corrin)]-$CN^+$. The proposed assignment includes only transitions that can be correlated with experimental bands. Note that for higher energy regions of the spectrum, several states can be associated with excitations observed in experiment. See the main text for further discussion and the Supporting Information for a full list of excitations (n denotes free electron pairs of corrin nitrogen atoms).

| | E(eV) | λ(nm) | *f* | % | Character | | | Experiment E(eV) λ(nm) |
|---|---|---|---|---|---|---|---|---|
| $S_1$ | 2.43 | 510.4 | 0.0207 | 30 | 117 → 120 | H-2 → L | $d_{xz}+\pi_{CN}+\pi \rightarrow \pi^*$ | 2.25 (550) |
| | | | | 60 | 119 → 120 | H → L | $\pi \rightarrow \pi^*$ | |
| $S_2$ | 2.53 | 490.8 | 0.0237 | 72 | 118 → 120 | H-1→ L | $d_{yz}+\pi \rightarrow \pi^*$ | |
| $S_3$ | 2.62 | 474.1 | 0.0435 | 56 | 117 → 120 | H-2 → L | $d_{xz}+\pi_{CN}+\pi \rightarrow \pi^*$ | 2.40 (517) |
| | | | | 18 | 119 → 120 | H → L | $\pi \rightarrow \pi^*$ | |
| | | | | 14 | 119 → 121 | H → L+1 | $\pi \rightarrow d_{xy}+n$ | |
| $S_4$ | 2.73 | 454.7 | 0.0137 | 74 | 119 → 121 | H → L+1 | $\pi \rightarrow d_{xy}+n$ | 2.56 (485) |
| $S_5$ | 2.85 | 434.5 | 0.0034 | 72 | 118 → 121 | H-1→ L+1 | $d_{yz}+\pi \rightarrow d_{xy}+n$ | |
| $S_6$ | 2.89 | 429.3 | 0.0039 | 84 | 116 → 120 | H-3 → L | $d_{x2-y2}+\pi_{CN} \rightarrow \pi^*$ | |
| $S_7$ | 3.00 | 413.7 | 0.0071 | 76 | 118 → 122 | H-1→ L+2 | $d_{yz}+\pi \rightarrow \sigma^*(d_{z2})+n$ | |
| $S_8$ | 3.00 | 413.0 | 0.0124 | 70 | 119 → 122 | H → L+2 | $\pi \rightarrow \sigma^*(d_{z2})+n$ | 2.71 (457) |
| | | | | 14 | 119 → 123 | H → L+3 | $\pi \rightarrow \pi^*$ | |
| $S_9$ | 3.11 | 399.1 | 0.0013 | 68 | 115 → 120 | H-4 → L | $d_{x2-y2}+\pi_{CN}+\pi \rightarrow \pi^*$ | |
| | | | | 18 | 119 → 123 | H → L+3 | $\pi \rightarrow \pi^*$ | |
| $S_{10}$ | 3.20 | 387.2 | 0.0091 | 10 | 113 → 120 | H-6 → L | $\pi_{CN}+\pi \rightarrow \pi^*$ | 3.04 (408) |
| | | | | 66 | 118 → 123 | H-1→ L+3 | $d_{yz}+\pi \rightarrow \pi^*$ | |
| . . . | | | | | | | | |
| $S_{15}$ | 3.54 | 350.2 | 0.1052 | 16 | 115 → 121 | H-4 → L+1 | $d_{x2-y2}+\pi_{CN}+\pi \rightarrow d_{xy}+n$ | |
| | | | | 30 | 119 → 123 | H → L+3 | $\pi \rightarrow \pi^*$ | |
| $S_{16}$ | 3.56 | 348.3 | 0.0399 | 60 | 113 → 120 | H-6 → L | $\pi_{CN}+\pi \rightarrow \pi^*$ | 3.44 (360) |
| | | | | 10 | 119 → 124 | H → L+4 | $\pi \rightarrow \pi^*$ | |
| . . . | | | | | | | | |
| $S_{23}$ | 3.91 | 316.8 | 0.0302 | 70 | 118 → 124 | H-1→ L+4 | $d_{yz}+\pi \rightarrow \pi^*$ | |
| $S_{24}$ | 3.92 | 316.0 | 0.0157 | 18 | 113 → 121 | H-6 → L+1 | $\pi_{CN}+\pi \rightarrow d_{xy}+n$ | 3.85 (322) |
| | | | | 26 | 115 → 123 | H-4 → L+3 | $d_{x2-y2}+\pi_{CN}+\pi \rightarrow \pi^*$ | |
| | | | | 28 | 119 → 124 | H → L+4 | $\pi \rightarrow \pi^*$ | |
| . . . | | | | | | | | |
| $S_{29}$ | 4.10 | 302.7 | 0.0150 | 30 | 111 → 120 | H-8 → L | $\pi_{CN}+\pi+d_{yz} \rightarrow \pi^*$ | |
| | | | | 12 | 113 → 122 | H-6 → L+2 | $\pi_{CN}+\pi \rightarrow \sigma^*(d_{z2})+n$ | |
| | | | | 12 | 117 → 124 | H-2 → L+4 | $d_{xz}+\pi_{CN}+\pi \rightarrow \pi^*$ | |
| | | | | 12 | 119 → 125 | H → L+5 | $\pi \rightarrow \pi^*_{Im}$ | |
| $S_{30}$ | 4.11 | 301.8 | 0.0129 | 34 | 117 → 124 | H-2 → L+4 | $d_{xz}+\pi_{CN}+\pi \rightarrow \pi^*$ | |
| | | | | 58 | 118 → 125 | H-1→ L+5 | $d_{yz}+\pi \rightarrow \pi^*_{Im}$ | |
| $S_{31}$ | 4.12 | 300.7 | 0.0261 | 46 | 117 → 124 | H-2 → L+4 | $d_{xz}+\pi_{CN}+\pi \rightarrow \pi^*$ | |
| | | | | 38 | 118 → 125 | H-1→ L+5 | $d_{yz}+\pi \rightarrow \pi^*_{Im}$ | 4.05 (306) |
| . . . | | | | | | | | |
| $S_{33}$ | 4.25 | 291.6 | 0.0146 | 68 | 112 → 121 | H-7 → L+1 | $\sigma(d_{z2}) \rightarrow d_{xy}+n$ | |
| | | | | 12 | 113 → 122 | H-6 → L+2 | $\pi_{CN}+\pi \rightarrow \sigma^*(d_{z2})+n$ | |
| $S_{34}$ | 4.28 | 289.5 | 0.0505 | 14 | 111 → 120 | H-8 → L | $\pi_{CN}+\pi+d_{yz} \rightarrow \pi^*$ | |
| | | | | 24 | 112 → 121 | H-7 → L+1 | $\sigma(d_{z2}) \rightarrow d_{xy}+n$ | |
| | | | | 26 | 113 → 122 | H-6 → L+2 | $\pi_{CN}+\pi \rightarrow \sigma^*(d_{z2})+n$ | |
| . . . | | | | | | | | |
| $S_{38}$ | 4.48 | 277.0 | 0.0330 | 16 | 113 → 123 | H-6 → L+3 | $\pi_{CN}+\pi \rightarrow \pi^*$ | |
| | | | | 22 | 117 → 125 | H-2 → L+5 | $d_{xz}+\pi_{CN}+\pi \rightarrow \pi^*_{Im}$ | |
| $S_{39}$ | 4.50 | 275.4 | 0.0343 | 32 | 113 → 123 | H-6 → L+3 | $\pi_{CN}+\pi \rightarrow \pi^*$ | |
| . . . | | | | | | | | |
| $S_{43}$ | 4.66 | 266.2 | 0.0307 | 42 | 108 $\rightarrow$ 120 | H-11 $\rightarrow$ L | $\pi_{Im}+\pi+n \rightarrow \pi^*$ | 4.70 (264) |
| | | | | 20 | 115 $\rightarrow$ 124 | H-4 $\rightarrow$ L+4 | $d_{x2-y2}+\pi_{CN}+\pi \rightarrow \pi^*$ | |
| $S_{44}$ | 4.68 | 264.8 | 0.0422 | 24 | 108 $\rightarrow$ 120 | H-11 $\rightarrow$ L | $\pi_{Im}+\pi+n \rightarrow \pi^*$ | |
| | | | | 10 | 109 $\rightarrow$ 120 | H-10 $\rightarrow$ L | $\pi+n \rightarrow \pi^*$ | |
| | | | | 26 | 110 $\rightarrow$ 121 | H-9 $\rightarrow$ L+1 | $\pi+\pi_{CN}+d_{xz} \rightarrow d_{xy}+n$ | |

**Table 5.** TD-DFT/BP86/6-31G(d)-based electronic transitions for the singlet states of Im-[CoIII(corrin)]-CN+ calculated with the use of COSMO ($H_2O$) solvation model. The proposed assignment includes only transitions that can be correlated with experimental bands. Note that for higher energy regions of the spectrum, several states can be associated with excitations observed in experiment. See the main text for further discussion and the Supporting Information for a full list of excitations.

| | E(eV) | λ(nm) | *f* | % | Character | | | Experiment E(eV) λ(nm) |
|---|---|---|---|---|---|---|---|---|
| $S_1$ | 2.47 | 502.2 | 0.0409 | 88 | 119 → 120 | H → L | $\pi \rightarrow \pi^*$ | 2.25 (550) |
| $S_2$ | 2.61 | 474.5 | 0.0155 | 50 | 118 → 120 | H-1 → L | $\pi+d_{yz} \rightarrow \pi^*$ | |
| | | | | 40 | 119 → 121 | H → L+1 | $\pi \rightarrow d_{xy}+n$ | |
| $S_3$ | 2.65 | 467.5 | 0.0136 | 53 | 119 → 121 | H → L+1 | $\pi \rightarrow d_{xy}+n$ | 2.40 (517) |
| | | | | 29 | 118 → 120 | H-1 → L | $\pi+d_{yz} \rightarrow \pi^*$ | |
| $S_4$ | 2.79 | 444.7 | 0.0267 | 75 | 117 → 120 | H-2 → L | $d_{xz}+\pi+\pi_{Im} \rightarrow \pi^*$ | 2.56 (485) |
| $S_5$ | 2.83 | 438.8 | 0.0121 | 83 | 119 → 122 | H → L+2 | $\pi \rightarrow \sigma^*(d_{z2})+n$ | |
| $S_6$ | 2.86 | 433.6 | 0.0039 | 78 | 118 → 121 | H-1 → L+1 | $\pi+d_{yz} \rightarrow d_{xy}+n$ | |
| $S_7$ | 2.94 | 421.3 | 0.0078 | 78 | 118 → 122 | H-1 → L+2 | $\pi+d_{yz} \rightarrow \sigma^*(d_{z2})+n$ | |
| | | | | 10 | 118 → 121 | H-1 → L+1 | $\pi+d_{yz} \rightarrow d_{xy}+n$ | 2.71 (457) |
| $S_8$ | 3.01 | 411.8 | 0.0002 | 89 | 116 → 120 | H-3 → L | $d_{x2-y2}+\pi_{Im} \rightarrow \pi^*$ | |
| $S_9$ | 3.05 | 406.8 | 0.0075 | 86 | 115 → 120 | H-4 → L | $\pi_{Im}+ d_{x2-y2} \rightarrow \pi^*$ | |
| $S_{10}$ | 3.30 | 375.4 | 0.0095 | 30 | 118 → 123 | H-1 → L+3 | $\pi+d_{yz} \rightarrow \pi^*$ | |
| | | | | 28 | 117 → 121 | H-2 → L+1 | $d_{xz}+\pi+\pi_{Im} \rightarrow d_{xy}+n$ | |
| | | | | 26 | 117 → 122 | H-2 → L+2 | $d_{xz}+\pi+\pi_{Im} \rightarrow \sigma^*(d_{z2})+n$ | |
| $S_{11}$ | 3.32 | 373.3 | 0.0095 | 49 | 118 → 123 | H-1 → L+3 | $\pi+d_{yz} \rightarrow \pi^*$ | 3.04 (408) |
| | | | | 18 | 117 → 122 | H-2 → L+2 | $d_{xz}+\pi+\pi_{Im} \rightarrow \sigma^*(d_{z2})+n$ | |
| | | | | 13 | 117 → 121 | H-2 → L+1 | $d_{xz}+\pi+\pi_{Im} \rightarrow d_{xy}+n$ | |
| $S_{12}$ | 3.33 | 372.5 | 0.0256 | 41 | 119 → 123 | H → L+3 | $\pi \rightarrow \pi^*$ | |
| | | | | 40 | 114 → 120 | H-5 → L | $d_{yz}+\pi+\pi_{CN} \rightarrow \pi^*$ | |
| . . . | | | | | | | | |
| $S_{17}$ | 3.68 | 337.4 | 0.1367 | 32 | 114 → 120 | H-5 → L | $d_{yz}+\pi+\pi_{CN} \rightarrow \pi^*$ | 3.44 (360) |
| | | | | 29 | 119 → 123 | H → L+3 | $\pi \rightarrow \pi^*$ | |
| | | | | 12 | 117 → 123 | H-2 → L+3 | $d_{xz}+\pi+\pi_{Im} \rightarrow \pi^*$ | |
| . . . | | | | | | | | |
| $S_{22}$ | 3.99 | 310.8 | 0.0162 | 33 | 118 → 124 | H-1 → L+4 | $\pi+d_{yz} \rightarrow \pi^*$ | |
| | | | | 26 | 114 → 122 | H-5 → L+2 | $d_{yz}+\pi+\pi_{CN} \rightarrow \sigma^*(d_{z2})+n$ | |
| $S_{23}$ | 4.01 | 309.6 | 0.0048 | 46 | 118 → 124 | H-1 → L+4 | $\pi+d_{yz} \rightarrow \pi^*$ | |
| | | | | 12 | 114 → 121 | H-5 → L+1 | $d_{yz}+\pi+\pi_{CN} \rightarrow d_{xy}+n$ | |
| $S_{24}$ | 4.08 | 304.2 | 0.0218 | 20 | 114 → 121 | H-5 → L+1 | $d_{yz}+\pi+\pi_{CN} \rightarrow d_{xy}+n$ | |
| | | | | 18 | 114 → 122 | H-5 → L+2 | $d_{yz}+\pi+\pi_{CN} \rightarrow \sigma^*(d_{z2})+n$ | 3.85 (322) |
| | | | | 15 | 114 → 123 | H-5 → L+3 | $d_{yz}+\pi+\pi_{CN} \rightarrow \pi^*$ | |
| $S_{25}$ | 4.15 | 298.7 | 0.0204 | 20 | 113 → 121 | H-6 → L+1 | $\pi+\pi_{CN}+d_{xz} \rightarrow d_{xy}+n$ | |
| | | | | 18 | 114 → 123 | H-5 → L+3 | $d_{yz}+\pi+\pi_{CN} \rightarrow \pi^*$ | |
| | | | | 13 | 114 → 121 | H-5 → L+1 | $d_{yz}+\pi+\pi_{CN} \rightarrow d_{xy}+n$ | |
| | | | | 12 | 119 → 124 | H → L+4 | $\pi \rightarrow \pi^*$ | |
| $S_{26}$ | 4.20 | 295.5 | 0.0092 | 64 | 113 → 121 | H-6 → L+1 | $\pi+\pi_{CN}+d_{xz} \rightarrow d_{xy}+n$ | |
| . . . | | | | | | | | |
| $S_{29}$ | 4.30 | 288.6 | 0.0176 | 85 | 117 → 124 | H-2 → L+4 | $d_{xz}+\pi+\pi_{Im} \rightarrow \pi^*$ | |
| $S_{30}$ | 4.32 | 286.8 | 0.0043 | 83 | 113 → 122 | H-6 → L+2 | $\pi+\pi_{CN}+d_{xz} \rightarrow \sigma^*(d_{z2})+n$ | |
| $S_{31}$ | 4.42 | 280.3 | 0.0079 | 78 | 119 → 125 | H → L+5 | $\pi \rightarrow \pi^*_{Im}$ | |
| | | | | 10 | 114 → 123 | H-5 → L+3 | $d_{yz}+\pi+\pi_{CN} \rightarrow \pi^*$ | |
| $S_{32}$ | 4.44 | 279.2 | 0.0291 | 81 | 111 → 120 | H-8 → L | $\pi_{CN}+\pi \rightarrow \pi^*$ | 4.05 (306) |
| $S_{33}$ | 4.50 | 275.3 | 0.0193 | 49 | 110 → 120 | H-9 → L | $\pi_{Im}+\pi_{CN}+\pi \rightarrow \pi^*$ | |
| | | | | 14 | 119 → 125 | H → L+5 | $\pi \rightarrow \pi^*_{Im}$ | |
| | | | | 13 | 116 → 124 | H-3 → L+4 | $d_{x2-y2}+\pi_{Im} \rightarrow \pi^*$ | |
| . . . | | | | | | | | |
| $S_{40}$ | 4.76 | 260.3 | 0.0780 | 27 | 112 → 121 | H-7 → L+1 | $\pi_{Im}+\pi_{CN} \rightarrow d_{xy}+n$ | 4.70 (264) |
| | | | | 19 | 111 → 121 | H-8 → L+1 | $\pi_{CN}+\pi \rightarrow d_{xy}+n$ | |

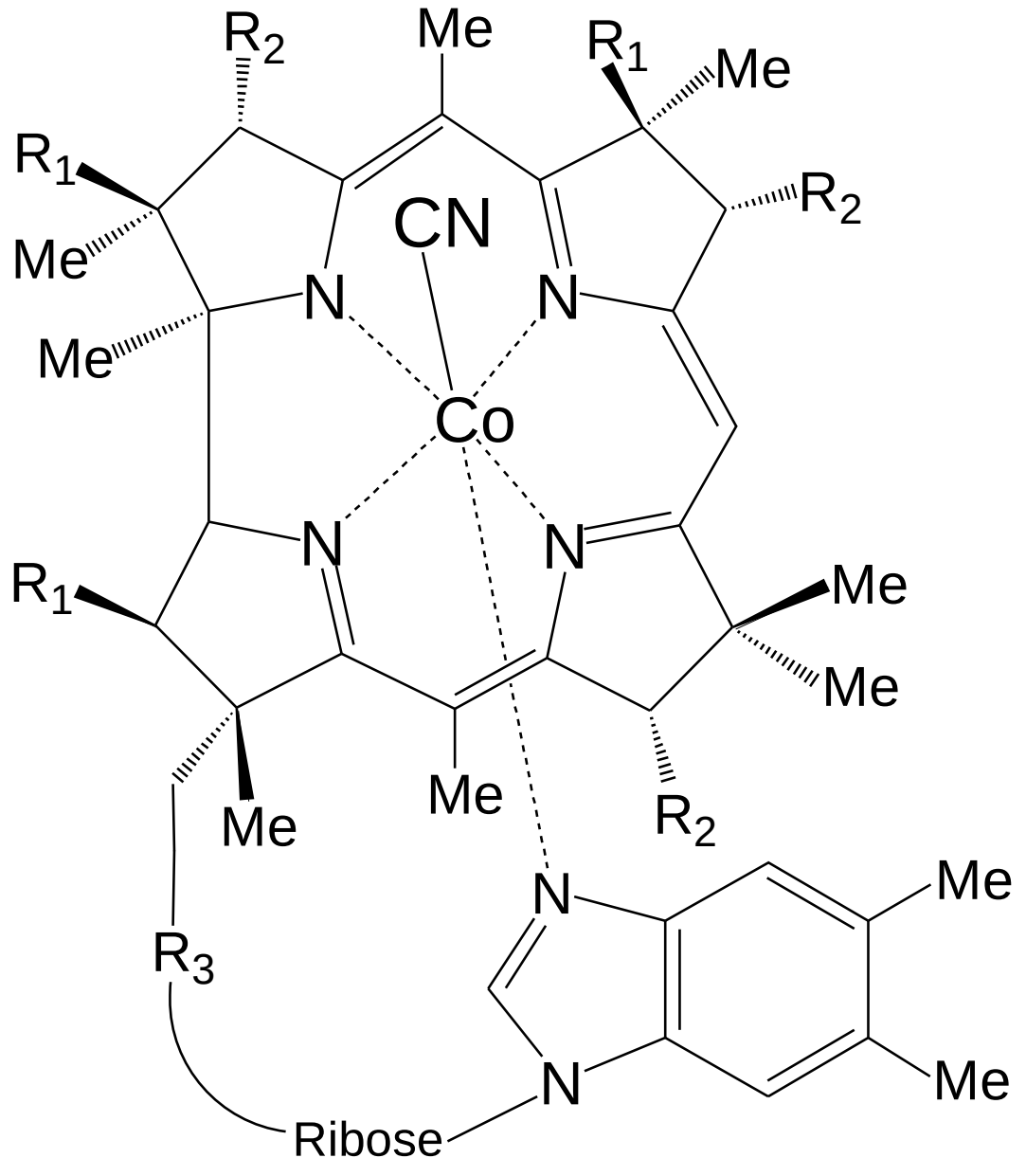

**Figure 1.** Molecular structure of vitamin $B_{12}$ (cyanocobalamin or CNCbl).

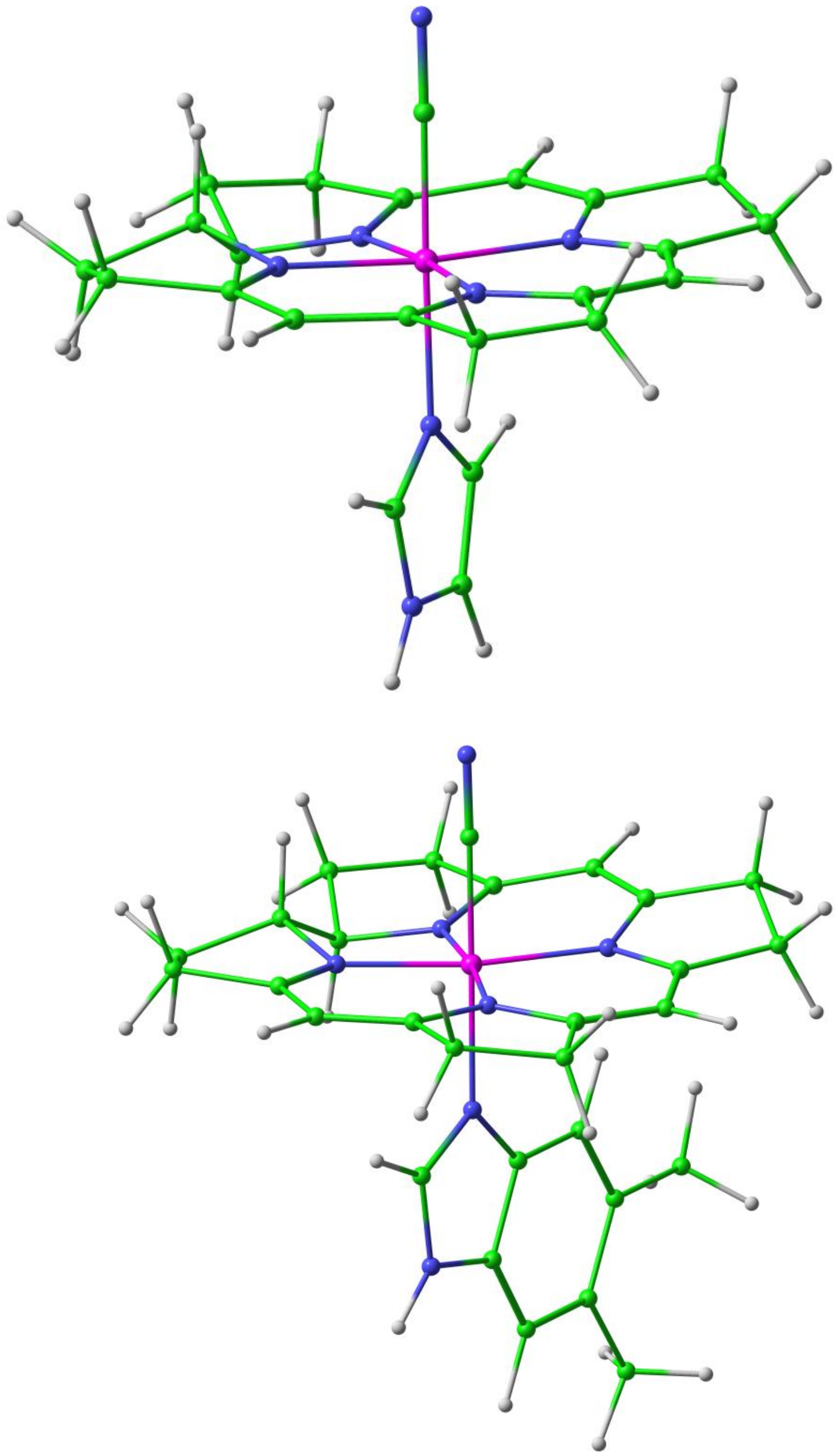

**Figure 2.** Two structural models used in excited state calculations (upper – with imidazole, bottom – with dimethylbenzimidazole as lower axial ligand).

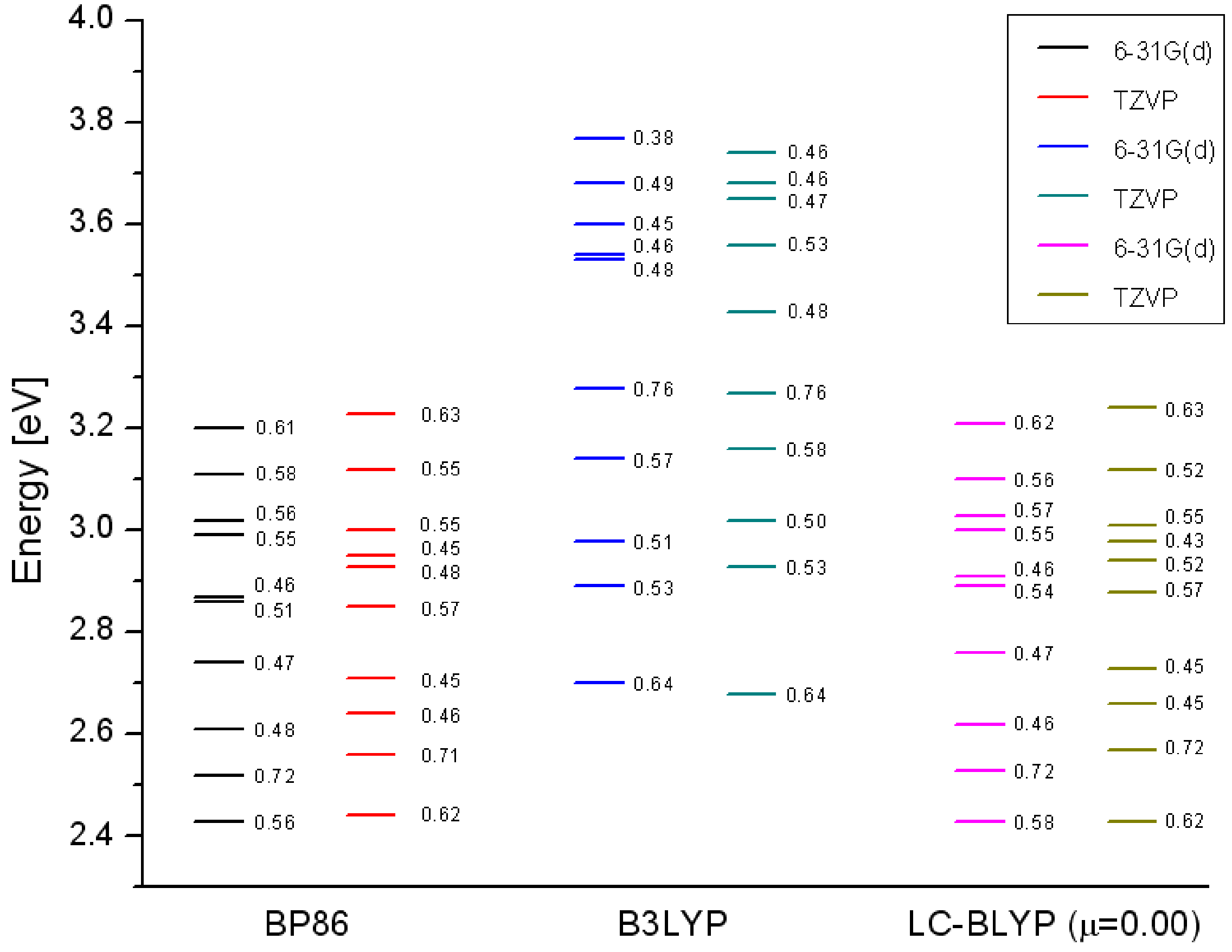

**Figure 3.** Ten lowest electronic transitions of Im-[Co$^{III}$(corrin)]-CN$^{+}$ calculated with different density functionals and basis sets together with the corresponding $\Lambda$ diagnostic values (presented as the numbers).

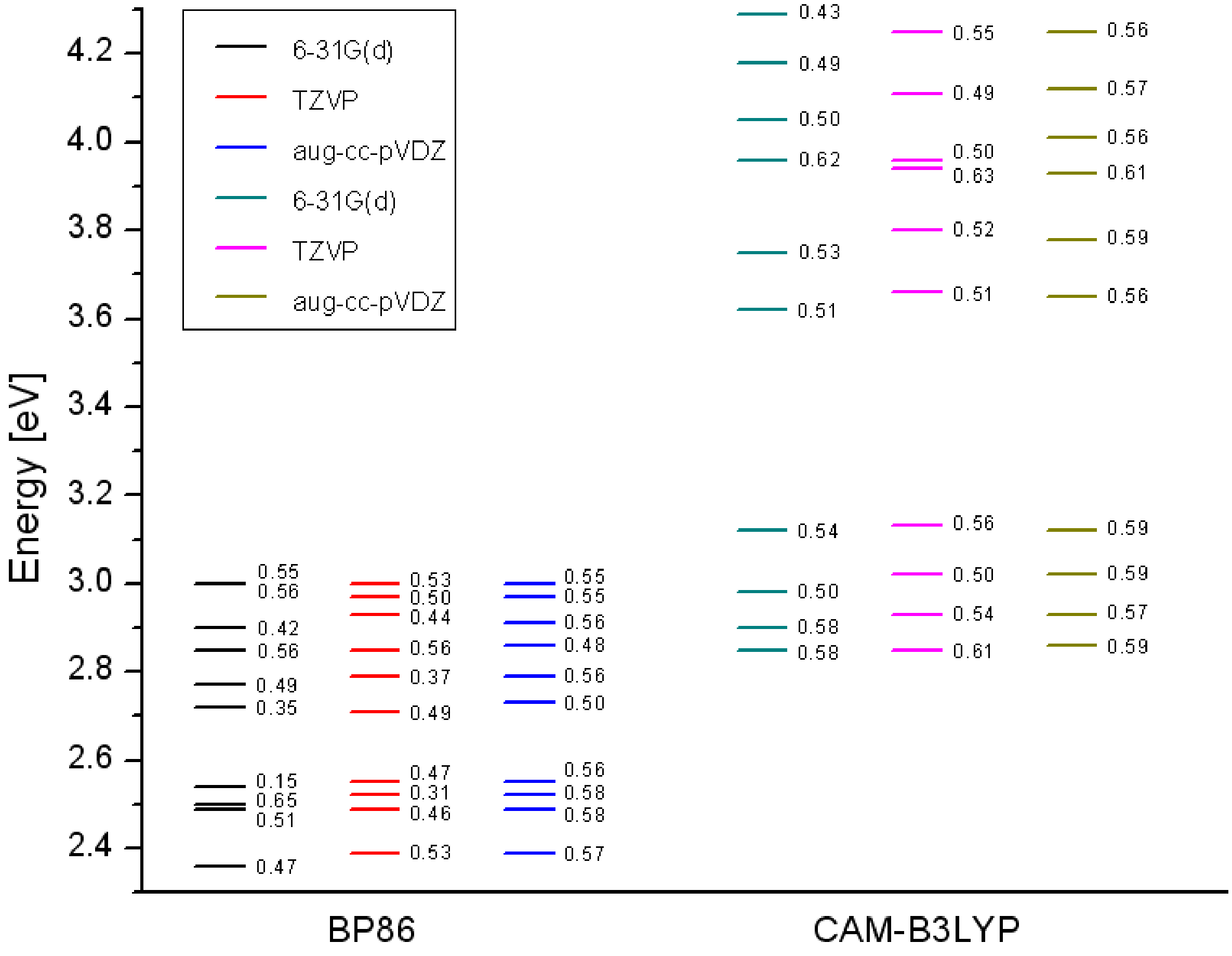

**Figure 4.** Ten lowest electronic transitions of DBI-[Co$^{III}$(corrin)]-CN$^+$ calculated with different density functionals and basis sets together with the corresponding $\Lambda$ diagnostic values (presented as numbers).

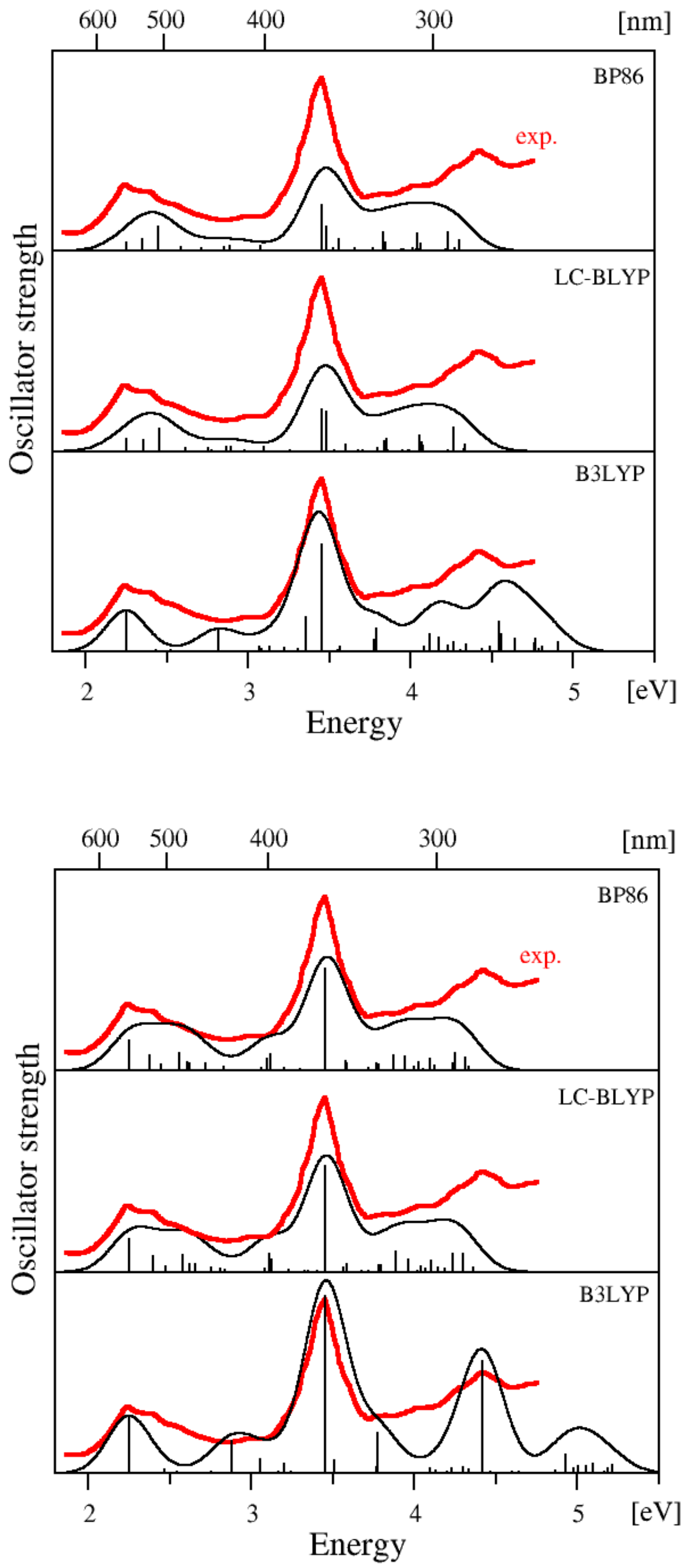

**Figure 5.** Simulated electronic absorption spectra of CNCbl computed with the use of different functionals and the 6-31G(d) basis set (upper – vacuo, bottom – water solution, PCM). See Tables: S3 and S4 (Supporting information) for details regarding the scaling factors. Experimental spectra were reproduced from Ref. 51 with permission.

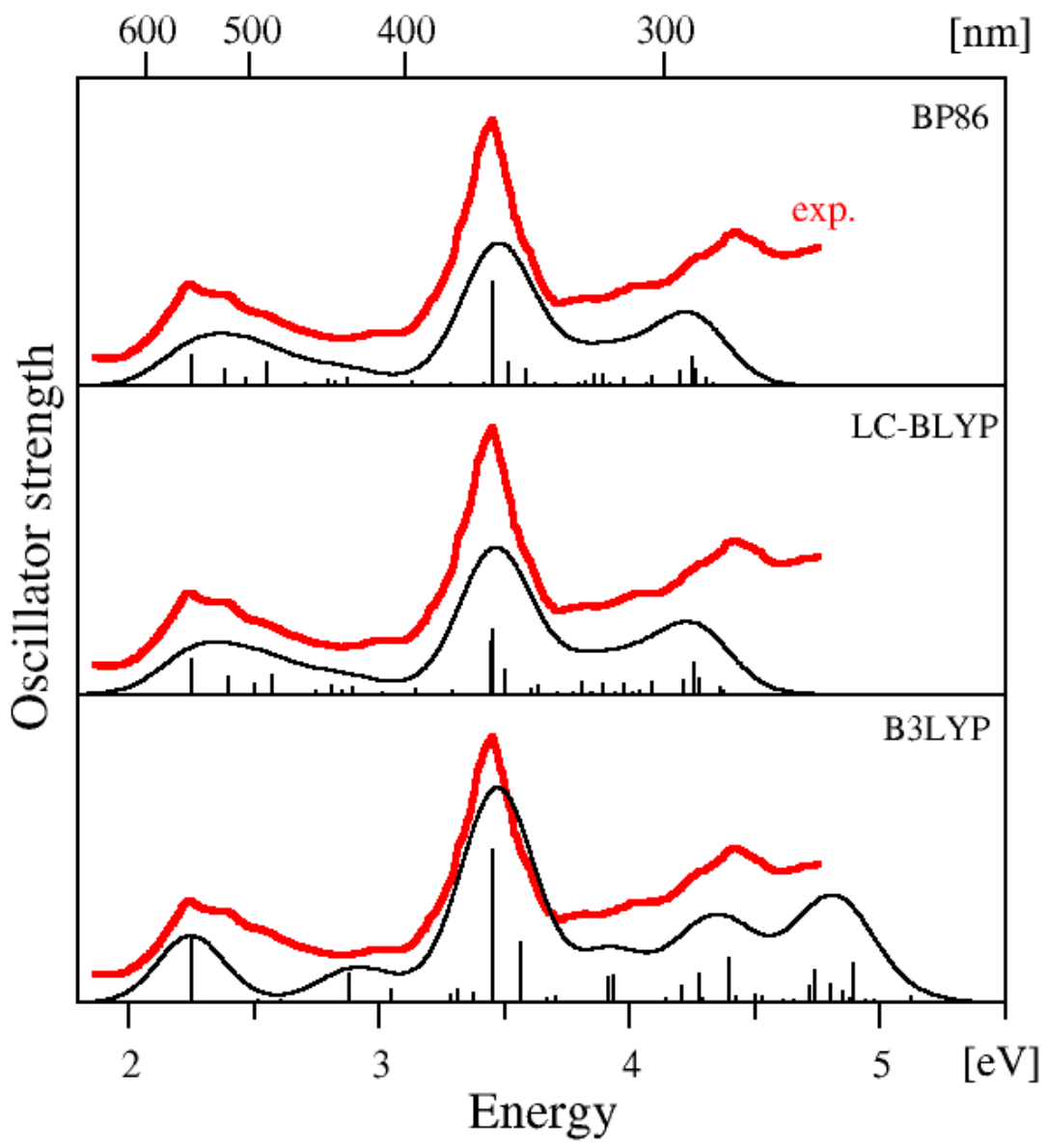

**Figure 6.** Simulated electronic absorption spectra of CNCbl computed with the use of different functionals and the TZVP basis set (vacuo). See Table S5 (Supporting information) for details regarding the scaling factors. Experimental spectra were reproduced from Ref. 51 with permission.

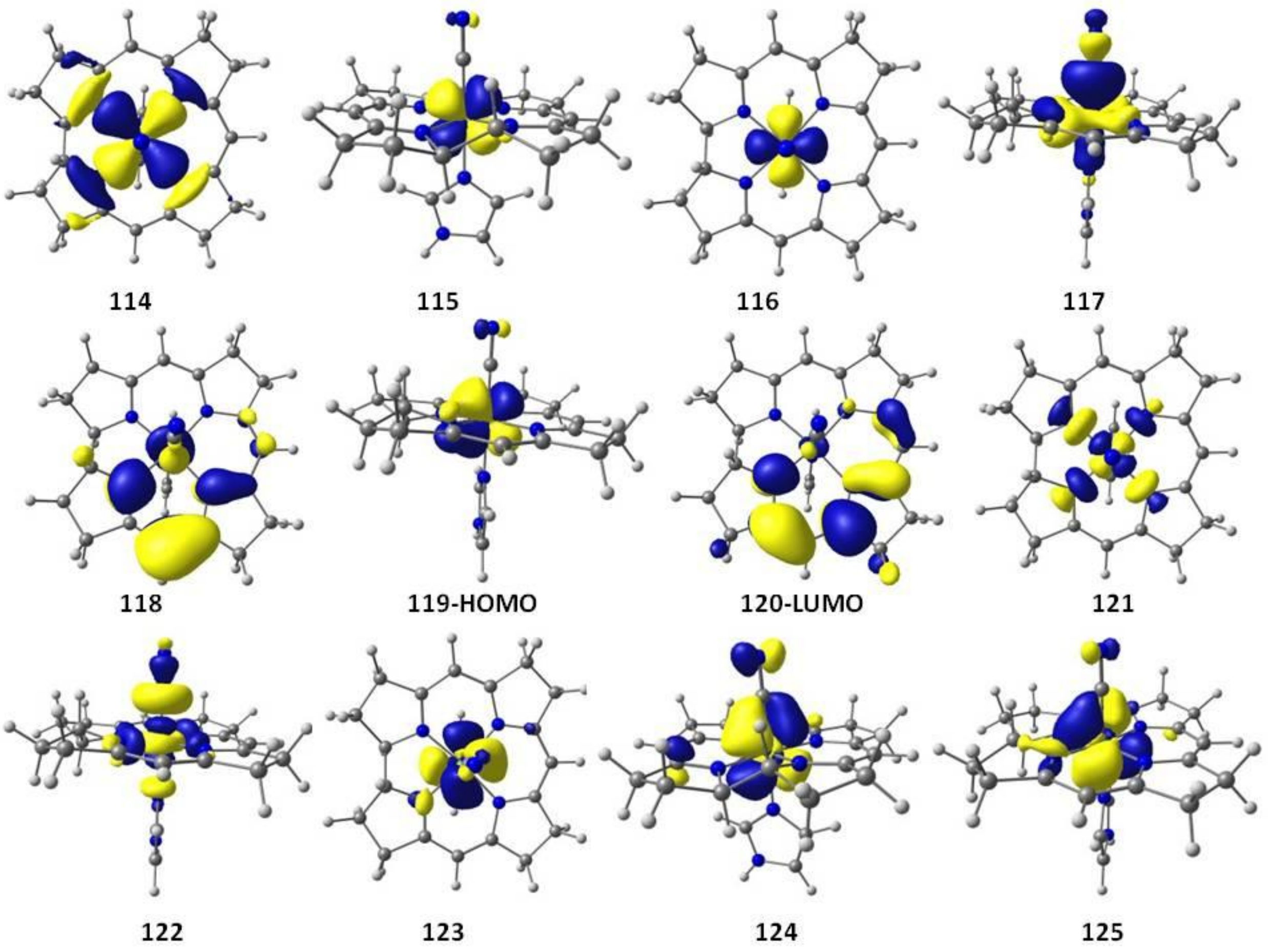

**Figure 7.** Active space orbitals of Im-[$Co^{III}$(corrin)]-$CN^+$ used in CASSCF(12,12) calculations.

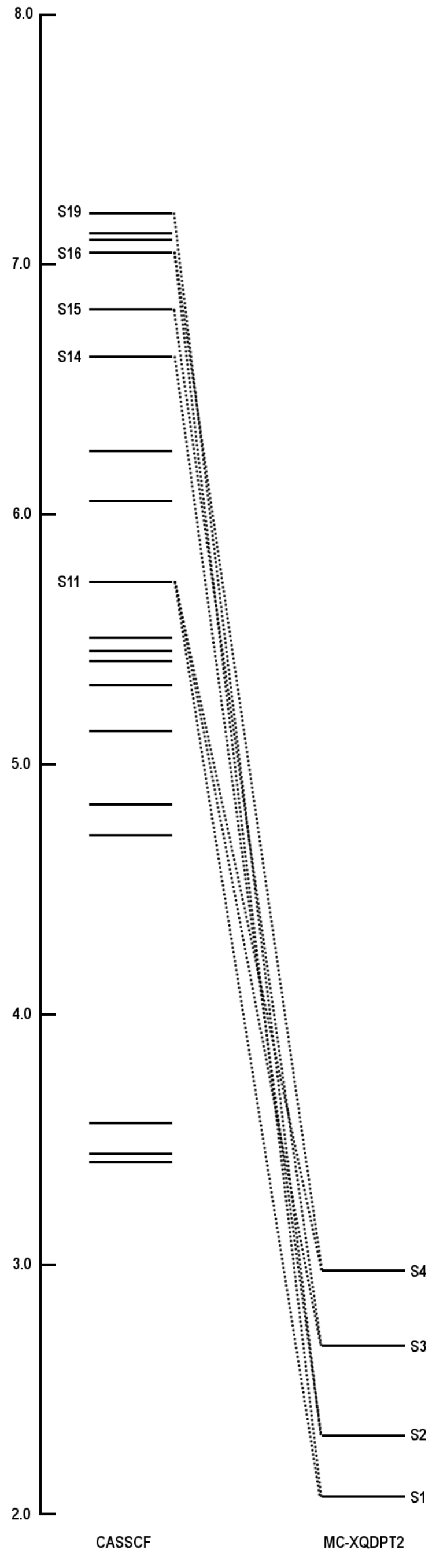

**Figure 8.** Energies of the lowest twenty excited states computed at the CASSCF level and the corresponding four low-lying states using MC-XQDPT2. The lines between CASSCF and MC-XQDPT2 indicate mixing of states upon inclusion of second order perturbation theory (see Table 2 for further details).

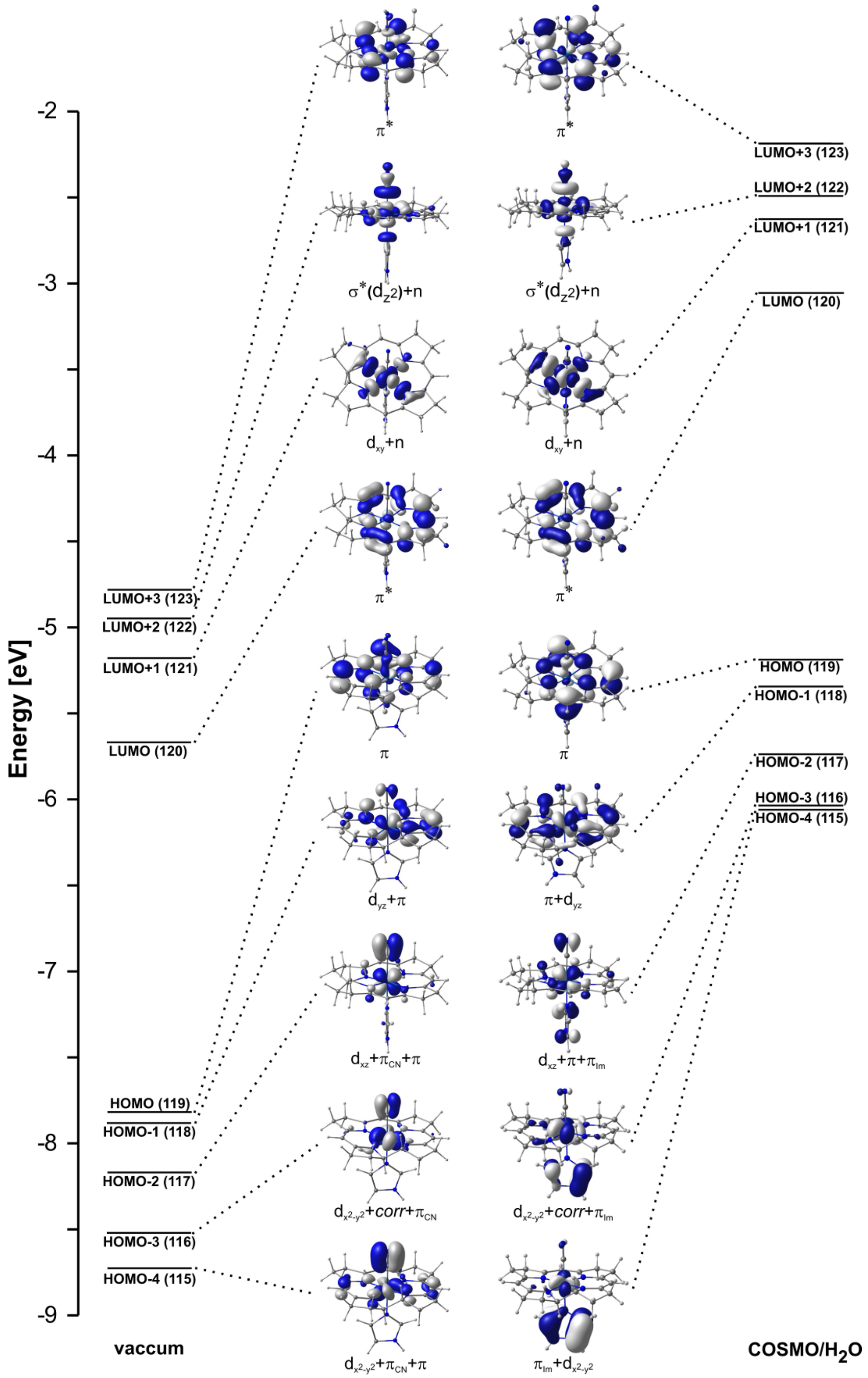

**Figure 9.** Isosurface plots of frontier MOs for the ground state of CNCbl model. Energies are based on BP86/6-31G(d) calculations in vacuo and with the COSMO/$H_2O$ model.

[58] Roos, B. O.; Veryazov, V.; Conradie, J.; Taylor, P. R.; Ghosh, A. *J. Phys. Chem. B* **2008**, *112*, 14099-14102.